\documentclass[a4paper,12pt]{article}%
\usepackage{amsfonts}
\usepackage{makeidx}
\usepackage{latexsym}
\usepackage{amsmath}
\usepackage{bbold}
\usepackage{color}
\usepackage{amssymb}
\usepackage{graphicx, amsmath, amsthm, amssymb, bm, boxedminipage}
\usepackage{graphicx}%
\allowdisplaybreaks[4]
\allowdisplaybreaks[4]

\numberwithin{equation}{section}
\theoremstyle{plain}
\newtheorem{thm}{Theorem}[section]
\newtheorem{lemma}[thm]{Lemma}

\newtheorem{corol}[thm]{Corollary}

\usepackage[english]{babel}
\usepackage[autostyle, english = american]{csquotes}
\MakeOuterQuote{"}

\usepackage{hyperref}
\hypersetup{
    colorlinks = true, 
    urlcolor = blue,   
    linkcolor = blue,  
    citecolor = red    
}

\begin{document}

\title{Price-Based Attention and Welfare\thanks{Department of Economics, McGill University,
kaushil.patel@mail.mcgill.ca. I am deeply
indebted to Rohan Dutta and Larry Epstein for their constant support and generous feedback throughout this research project and my PhD. I thank also participants at the CIREQ lunch seminar at McGill University for their helpful comments.}}
\author{Kaushil Patel}

\maketitle
\centerline{Please \href{https://drive.google.com/file/d/1mCapxKGlCpr6Zs4zQZRn_c5DWD-EkVqZ/view?usp=drive_link}{click here for the latest draft}.} 

\begin{abstract}
To choose between two discrete goods, a consumer pays attention to only those with prices below a threshold. From these, she chooses her most preferred good. We assume consumers in a population have the same preference but may have different thresholds. Similar models of bounded rationality have been studied in the empirical marketing literature. We fully characterize the model, and using observational choice data alone, we identify the welfare implications of a price change. The behavioral content of our model overlaps with an important class of random utility models, but the welfare implications are meaningfully different. The distribution of equivalent variation under our model first-order stochastically dominates that under the random utility model.


\medskip

\bigskip


\end{abstract}

\newpage

\section{Introduction}
\subsection{Objectives and Outline}
Classical models of discrete choice assume that consumers are rational and differences in choices arise from unobservable differences in preferences. Evidence suggests that another unobservable source of heterogeneity may be attention. All consumers may not consider the same goods even if the same goods are available to all of them. Models of bounded rationality take limited attention explicitly into account. One can ask a few natural questions: What is the empirical content of a model with limited attention consumers? Is welfare analysis possible, and if so, under what conditions? Is a population of boundedly rational consumers with the same preference distinguishable from a population of rational consumers with heterogeneous preferences? This paper addresses these questions in a binary choice setting for a well-known model of price-based attention.

Consider a population of decision makers (DMs) with identical income who follow a two-step procedure when choosing from a binary set of available goods. Step one: each DM considers only those alternatives with a price less than a DM-specific threshold price. We refer to this price as her \textit{attention-price threshold} (or \textit{attention threshold}). Step two: she picks the utility-maximizing alternative from the alternatives she considers. We assume that the utility function in step two is common across all agents.

Attention-price thresholds as a step to eliminate alternatives from consideration were first proposed in psychology by Tversky (1972b). Decision makers eliminate alternatives by aspects, where price may be an aspect, until only one alternative remains. In economics, Manzini and Mariotti (2012) study a model where in the first step, the consumer categorizes the available alternatives and only considers alternatives from an undominated category. In the second step, she picks the best one according to her preference. They suggest the use of prices to form categories but focus on the more general model for analysis. We provide an axiomatization for attention-price thresholds directly. 

The use of attention-price thresholds to eliminate alternatives has been validated empirically in the marketing literature. Yee et al. (2007) find in survey experiments with students that for half of the respondents a high price affected which smartphones they considered and price was the first aspect that they focused on. Price was the only aspect (out of 16 aspects of smartphones) that was used to eliminate alternatives while other aspects were used to “accept.” More recently, Jagabathula and Rusmevichientong (2017) find that a model with attention-price thresholds and random preferences significantly improves upon the multinomial logit in predicting market shares of consumer goods at grocery stores. The key differences between their model and ours are that we assume a common preference in the population and study binary choice. Their main results are an estimation algorithm to fit the model to data and an algorithm to solve the firm's joint assortment and price choice problem. On the other hand, we provide a behavioral characterization for the model with a common preference, and identify welfare change. The two papers are thus complementary. 

This paper contributes to both theoretical and applied literatures on bounded rationality. Decision-theoretic models assume consumers choose from a menu of alternatives and describe the choice process through which an alternative is selected. Limited attention is often modeled with a consideration set, which is a subset of the menu (Manzini and Mariotti 2014). These models typically fix a preference ranking and study the behavioral implication of various forms of attention heterogeneity. Analysis rests on observing choices from each possible menu, known as "menu variation" (Strzalecki 2025). It is often the case that models of limited attention can overlap with random utility models (RUMs) in the choice data they can rationalize (Cattaneo et al. 2020). An important insight of this literature is that standard revealed preference arguments break down when consumers have limited attention, and welfare analysis is complicated (Masatlioglu et al. 2012). 



The applied literature also studies limited attention, but in a different way.\footnote{Strzalecki 2025 comments that "...the two literatures don't talk to each other as much as they might want to."} Much of the focus is on consumer demand where goods possess a set of attributes. Each good is always available to the consumer but underlying attributes undergo variation ("attribute variation"). These models study preference and attention heterogeneity simultaneously. Unlike the theory literature, the focus is not on the empirical content of various forms of bounded rationality but rather on identifying preferences and attention when both vary. Identification typically requires auxiliary data, such as surveys about brand awareness (Draganska and Klapper 2011), or exclusion restrictions, which impose that certain attributes enter either utility or consideration but not both (Goeree 2008), or experimentation, which involves changing the choice environment so that consumers are observed making rational choices (Chetty et al. 2009, Taubinsky and Rees-Jones 2018). 


 

We make three contributions. First, we axiomatize the attention-price threshold (APT) model using price and income (attributes) variation in a binary choice setting. We show that welfare analysis is possible with observational choice data without requiring exclusion restrictions or auxiliary data, analogously to welfare analysis in the random utility model (RUM). Finally, we show that a population of rational consumers with quasi-linear random utility is indistinguishable from a population of consumers with attention-price thresholds and a common preference. 

Consider a binary choice problem (as in Taubinsky and Rees-Jones 2018, Bhattacharya 2021, and Coen 2023) with goods $0$ and $1$ available to all consumers with a common preference and income. Good $0$ can be viewed as an outside option with a price of zero that is considered by all consumers. Consumers choose one of the two goods to consume and spend their remaining income on a numeraire good. The data consists of the choice probability of good $1$ with rich variation in both income and prices. We show that the model is characterized by a set of five axioms. 

Next, we show that when the price of good $1$ increases, the distribution of the welfare change can be identified from observational choice data. If for all prices there exist consumers who pay full attention, then with enough variation in the price of good $1$ (holding income constant), the analyst can identify the distribution of equivalent variation (EV). If full attention consumers do not exist, then we can partially identify the distribution of EV. The key step in the result rests on observing the minimum price at which no one chooses good $1$. This price corresponds to the reservation price of rational consumers, and given the common preference, it is the price at which all consumers are indifferent between the two goods. Importantly, price affects both utility and attention and thus does not satisfy an exclusion restriction; nevertheless, price variation is sufficient to identify welfare. 

At the technical level, our welfare result builds on Bhattacharya (2015), who provides non-parametric welfare identification for RUM with unrestricted preference heterogeneity. We show that welfare analysis can be extended to models with limited attention, in a way analogous to how revealed preference theory has been extended to models of bounded rationality in the theory literature. The type of data required for welfare identification in the APT model is similar to that needed in random utility models: variation only in the price of good $1$.

The random utility model is the workhorse model for demand estimation and welfare analysis in discrete choice econometrics, and thus of particular interest for comparison with the attention-price threshold model. The key distinction between the RUM and APT models is that RUM assumes preference heterogeneity in the population and full attention while the APT model assumes a common preference and partial attention that is heterogeneous across individuals. Bhattacharya (2021) provides an axiomatization for RUM with unrestricted preference heterogeneity in binary choice. Our axiomatization of the APT model shows that it restricts income effects compared to RUM but allows discontinuities in the choice probability function, which are ruled out by RUM. 

For further comparison, we consider the special case of quasi-linear random utility (QRUM), where the utility functions are quasi-linear in price. These models are ubiquitous in the empirical literature on demand estimation and are good approximations when prices are small relative to income (McFadden 1973, Berry 1994, Berry and Pakes 2007, Train 2009, Dubé et al. 2020).\footnote{The canonical conditional logit of McFadden 1973 is a special case of the quasi-linear RUM where utility is linear in all attributes.} The main characteristic of QRUM is that it does not allow any income effects. We show that any choice data that can be rationalized by a QRUM can also be rationalized by an APT model. This is despite the source of heterogeneity being completely different. Real-world consumers likely differ in both their preferences and attention. Our comparison of the QRUM and APT models suggests that differentiating between the two sources of choice heterogeneity will be difficult in many empirically significant settings. However, the analyst can compare the welfare implications of the two models when they overlap. We show that the distribution of EV identified in the APT model first-order stochastically dominates the distribution of EV identified in RUM. In other words, RUM underestimates the welfare losses from price increases compared to the APT model. 

The rest of the paper is as follows. Section $1.2$ expounds on more related literature. Section $2$ presents the model. Section $3$ provides the empirical content of the attention-price threshold model. Section $4$ shows that welfare is identified. Section $5$ compares the axiomatization and identification results of the APT model with RUM, in particular the quasi-linear RUM. 



\subsection{More Related Literature}

There exists a large decision theory literature on models of bounded rationality (see Strzalecki 2025 or de Clippel and Rozen 2024 for surveys). Sequential choice procedures, like ours, were first characterized by Manzini and Mariotti (2007). A finding from this literature is that several models of boundedly rational choice are indistinguishable from models of rational choice. For example, Tversky's Elimination by Aspects is a special case of RUM (Tversky 1972a), and RUM is a special case of the random attention model of Cattaneo et al. (2020), which is a model with a fixed preference. We show that such results also hold for our model with attribute variation, which is generally not the case. 

A second finding from the theory literature concerns welfare analysis. Naive applications of revealed preference may be wrong because the direct link between preference and choice imposed by rationality is broken. Bernheim and Rangel (2009) suggest a "model-free" approach for behavioral welfare analysis. They propose an acyclic welfare criterion where an alternative $x$ is said to be unambiguously chosen over alternative $y$ if and only if $y$ is never chosen in any choice instance where $x$ is available. Given the strong requirement, the welfare criterion can be incomplete. Even when applicable, Masatlioglu et al. (2012) show that an analyst using the criterion would infer the decision maker's preference incorrectly in their model of limited attention. However, the correct model-based revealed preference relation in Masatlioglu et al. (2012) may also be incomplete, even with rich menu variation. Our approach is a model-based approach using attribute variation, and we show that welfare is identified. 

There also exists a large applied literature on limited attention with attribute variation. As discussed earlier, these papers typically assume both random preferences and random attention, and they focus on identification and estimation (see Crawford et al. 2021 for a review). Much of the literature requires auxiliary data and exclusion restrictions, e.g., prices affect only utility and not consideration. 

An exception to this approach are Abaluck and Adams-Prassl (2021), who show that when random utility is quasi-linear in prices and attention heterogeneity is given by two specific models of consideration set formation, then preference heterogeneity and attention heterogeneity can both be identified by price variation. Their result does not apply to our model for two reasons. The first is that with an outside option there need to be at least three alternatives for their identification theorems to apply, since they rely on cross-price derivatives of the choice probability function. Outside options are present in virtually all applied models of discrete choice (Berry and Haile 2021) and even in much of the theory literature. For applied work, the outside option is needed to ensure that the market demand elasticity is not zero.\footnote{Berry (1994) explain that without an outside good "...a general increase in prices will not decrease aggregate output; this is an unfortunate feature of some discrete choice models that have been applied to the empirical study of differentiated products markets..."} A second reason their identification results do not apply is that the APT model is not a special case of the two models of consideration sets they study. Thus, even in a multinomial choice setting, their results do not directly apply to the APT model.\footnote{It is not clear whether some modification of their argument would still apply in multinomial choice, but it is worth investigating.} Nevertheless, we show that welfare is identified in the binary choice APT model using price variation.

A strand of the empirical welfare analysis literature, especially in behavioral public finance, takes an experimental path by building on the "model-free" approach of Bernheim and Rangel (2009), further elaborated by Bernheim (2016). The core idea is that the analyst observes which choices are "optimal" and can use this knowledge to conduct welfare analysis when choices are "noisy," i.e., arise from bounded rationality.\footnote{This approach also exists in psychology by the name of "debiasing" (Soll et al. 2015)} Chetty et al. (2009) study the salience of sales taxes for consumers. They conduct an experiment where in one setting sales taxes are only visible at the register and in another setting, sales taxes are included in the price of the good. They find that demand falls significantly when taxes are posted in the price. They follow the Bernheim and Rangel (2009) approach by assuming that when taxes are included in the price, consumers are utility-maximizing. 


Taubinsky and Rees-Jones (2018) and Coen (2023) are two recent papers that follow the experimental approach to study the salience of taxes in a binary choice setting. Bhattacharya (2024) provides a short exposition on nonparametric empirical approaches to behavioral welfare analysis. Our model contributes to this literature by showing that welfare can be identified with only the "noisy" (observational) choice data for a certain form of bounded rationality. So, a finer understanding of the nature of bounded rationality can better guide empirical welfare analysis.

\section{Model}
Let there be two alternatives: $0$ and $1$. The price of good $0$ is $p_{0}=0$, and the price of good $1$ is $p_{1}\geq0$. 
For example, consider a school choice setting where the public school is good $0$ and a private school is good $1$. More generally, good $0$ can simply be interpreted as not choosing among the available consumer goods. There is a population of decision makers (DMs) with an individual denoted by $i$.\footnote{The population can be finite or infinite, where $i$ varies over an interval.} Each decision maker picks one good from those available to her and spends her remaining income (residual income) on a numeraire good. The budget constraint is given by
\[
 p_{1} \mathbb{1}\{\text{good 1 chosen}\}+z = y
\]

\noindent where $y>0$ is the DM's income, $z\in[0,y]$ is the quantity of numeraire chosen, and the indicator function determines which good is chosen. Income $y$ is homogeneous in the population. The quantity of numeraire consumed when good $0$ is chosen is $z=y$ while when good $1$ is chosen, the quantity of numeraire is $z=y-p_{1}$.  

Decision makers have a common preference. Let $U_{0}(y):(0,\infty)\xrightarrow{} [0,\infty)$ be the utility from consuming good $0$, and $U_{1}(y-p_{1}):[0,y]\xrightarrow{} [0,\infty)$ be the utility from consuming good $1$. Assumption 1 states two properties of the preference. First, the utility of good $0$ is non-decreasing in the numeraire, and the utility of good $1$ is continuous and strictly increasing in the numeraire. Second, for any income $y$, there exists a price high enough such that good $1$ is not the most preferred good. \\

\noindent\textbf{Assumption $1$ (Preferences)}: $(i) \ U_{0}(z)$ is non-decreasing in $z$, and\\ $U_{1}(z)$ is continuous and increasing in $z$.
\\
\indent$(ii)$ For any $y>0$, there exists a price $\bar{p_{1}}\in [0,y]$ such that $U_{0}(y)\geq$ \indent \indent $ U_{1}(y-\bar{p_{1}})$.\\ 

\noindent Assumption $1(ii)$ implies that when $p_{1}=y$, $U_{0}(y)\geq U_{1}(0) $. If good $1$ costs the consumer her entire income, then good $0$ is preferred to good 1. 

Decision makers follow a two-step procedure to choose an alternative. In the first step, decision maker $i$ considers those alternatives with prices less than ${t_{i}}$. We call ${t_{i}}$ the \textit{attention-price threshold} for decision maker $i$, and the set of alternatives she considers her consideration set. In the second step, she picks the utility-maximizing alternative from her consideration set. 

Attention may vary among the consumers. Let $G(t)$ be the cumulative distribution function (CDF) of attention-price thresholds, $t$, in the population. We interpret $G(\cdot)$
as the objective but unknown distribution of attention thresholds. If $G(t)=\alpha$, then the proportion of the population with attention-price thresholds less than or equal to $t$ is $\alpha$.

We make the following assumption regarding the consumers' attention. Assumption $2$ states that everyone pays attention to free goods (anyone who does not pay attention has mass zero). Consequently, good $0$ is considered by all consumers.  \\

\noindent \textbf{Assumption 2 (Full attention to free goods)}:  $G(0)=0$. \\


Denote by $q_{1}(p_{1},y)$ the probability that alternative $1$ is chosen when the price is $p_{1}$ and the income level is $y$. The \textit{model choice probability of alternative 1} is given by
\[
q_{1}(p_{1},y)= \mathbb{1}\{U_{0}(y)<U_{1}(y-p_{1})\} \,(1-G(p_{1})) \label{eq31} \text{.}
\]
We break utility-ties in favor of good $0$. Next, we define rationalization under the \textit{Attention-Price Threshold (APT) model}. \\

\noindent\textbf{Definition}: A choice probability function $q_{1}(p_{1},y)$ is  \textit{APT-rationalizable} if there exists a pair of utility functions $U_{0}(y)$ and $U_{1}(y-p)$ satisfying Assumption 1, and a distribution of attention $G(\cdot)$ satisfying Assumption $2$ such that
\begin{equation}
q_{1}(p_{1},y)= \mathbb{1}\{U_{0}(y)<U_{1}(y-p_{1})\} \,(1-G(p_{1})) \label{eq31} \text{.}
\end{equation}
\section{\label{Sec-Characterization}Characterization} 
In this section, we provide the empirical content of the APT model as the main theorem. We present the relevant axioms on choice behavior below. \\

\noindent\textbf{Axiom A}: $(i) \ q_{1}(p_{1}+\varepsilon,y+\varepsilon)$ is non-increasing in $\varepsilon$ for all $\varepsilon>0$. \newline \indent \indent \indent \vspace{1 mm} $(ii)\ q_{1}(\cdot,y)$ is non-increasing. \\

Axiom A$(i)$ states that if the price of good $1$ and income increase by the same amount, then the choice probability of good $1$ cannot increase. Alternatively, it says that if the residual income from consuming good $1$ stays the same while the residual income from good $0$ increases, then the choice probability of good $1$ cannot increase.\\

\noindent\textbf{Axiom B}: For any $y, y^{\prime}$ and $p_{1}$ , if $q_{1}(p_{1},y)>0$ and $q_{1}(p_{1},y^{\prime})>0$, then 
\[
q_{1}(p_{1},y)=q_{1}(p_{1},y^{\prime}) \text{.}
\]

\indent Axiom B states that if the choice probabilities of good $1$ are positive for the same price and two different levels of income, then the choice probabilities must be the same. Different levels of income affect whether consumers choose good $1$ at all. If consumers do choose good $1$, then they must prefer good $1$ to good $0$; consequently, the choice probability comes solely from the attention term, which depends only on the price. This implies strong restrictions on income effects. Specifically, if $q_{1}(p_{1},y)>0 $, then $q_{1}(p_{1},y^{\prime})=q_{1}(p_{1},y)$ or $q_{1}(p_{1},y^{\prime})=0$ for all $y^{\prime}>y$. The choice probability need not be monotonic in income and can alternate between $q_{1}(p_{1},y)$ and zero at higher levels of income.\\

\noindent\textbf{Axiom C}: For each $y$, there exists a $\bar{p_{1}}\in [0,y]$ such that $q_{1}(\bar{p_{1}},y)=0$. \\ 

Axiom C states that for any given income, there exists a price high enough such that no one buys the good. Combined with Axiom A$(ii)$, this implies that when price equals income no one buys good $1$.\\ 

\noindent\textbf{Axiom D}: For any $y$, if $q_{1}(p_{1}^{\prime},y)=0$ for all $p_{1}^{\prime}>p_{1}$, then $q_{1}(p_{1},y)=0$. \\

For a fixed income, if no one chooses good $1$ at any price higher than $p_{1}$, then no one chooses good $1$ at $p_{1}$ either. \\

\noindent\textbf{Axiom E}: For any $y$, $q_{1}(0,y)=0$ or $1$. \\

For any income, if the price of good $1$ is zero, then either no one buys the good or everyone buys the good. When the price is zero, everyone pays attention to the good. Given the common preference, either everyone buys or no one buys. 

\begin{thm}
    \label{Axiomatization} A choice probability function $q_{1}(\cdot,\cdot)$ is APT-rationalizable if and only if it satisfies Axioms A, B, C, D, and E. 
    \end{thm}
We provide a proof of Theorem \ref{Axiomatization} in the Appendix. Here, we provide a brief outline of the sufficiency part of the proof. We need to construct the three functions $U_{0}(\cdot)$, $U_{1}(\cdot)$, and $G(\cdot)$ satisfying Assumptions $1$ and $2$, respectively. Let  
\[
    U_{0}(y):=\max\{y-p_{1}:q_{1}(p_{1},y)=0\} \ \text{for all} \ y \text{.}
\]
The utility of good zero for any income level is set to the maximum residual income such that no one chooses good $1$. Let 
\[
    U_{1}(y-p_{1}):=y-p_{1} \textbf{.}
\] The utility of good $1$ for any residual income equals the residual income. Finally, let 
\[
G(p_{1}):=1-q_{1}(p_{1},y) \text{ for } q_{1}(p_{1},y)>0 \text{.}
\]
Whenever the choice probability at some price is positive, we set the proportion of consumers with attention thresholds less than or equal to that price equal to the proportion of consumers who did not consume the good. One can then check that these functions do satisfy the rationalizability condition (\ref{eq31}). 

The construction of $G(\cdot)$ reveals that for any price such that the choice probability is positive at some income, attention is identified. The proportion of consumers who pay attention to good $1$ equals the proportion who buy the good. 

Finally, if the analyst observes covariates, then these can be incorporated into the analysis similar to RUM. The APT model assumes that conditional on observables consumers have the same preference, and any heterogeneity in choices arises from heterogeneity in attention.

\section{\label{Sec-Welfare}Welfare Identification}
The aim of this section is to identify the distribution of welfare change when the price of a good changes. We will use equivalent variation (EV) as our money-metric measure of the change in welfare. The equivalent variation of a price increase is the amount of income that must be taken away from the decision maker at the original prices so that her utility from her choice is equal to her utility from her choice at the new prices. We formally state the definition in terms of this model in Appendix 5.1.

Consider an increase in the price of good 1 from $p_{1}^{\prime}$ to $p_{1}^{\prime\prime}$. Assume the price of good $0$ is fixed at $p_{0}=0$, and each member of the population has an income of $y$. We adopt a stronger version of Assumption $1$. Previously, we had assumed $U_{0}(z)$ was non-decreasing in $z$. We strengthen the monotonicity assumption by assuming $U_{0}(\cdot)$ is increasing. This ensures that for each consumer there is a unique value of equivalent variation. 

Adopt Assumption $2$ on attention as before. Consider the following additional assumption on attention that is important for welfare identification. \\

\noindent \textbf{Assumption $3$ (Positive probability of full attention)}: $G(t)<1$ for all $t<\infty$. \\

Assumption $3$ states that the support of $G(\cdot)$ has no maximum. For a finite population, let $g(t)$ be the probability mass function and assume $g(+\infty)>0$. Assumption $3$ states that for any price, there exist consumers with attention thresholds greater than that price. Thus, for any price of good $1$, there are always consumers who choose rationally.

We did not impose Assumption $3$ in our characterization, and it may not be satisfied for certain APT-rationalizable choice probability functions. Consider a choice probability function that is continuous at the minimum price such that no one chooses good $1$. By Axiom C, we know that for any income, there exists a price such that no one chooses good $1$. By Axiom D and monotonicity (Axiom A), we know that the minimum exists. Then, all attention thresholds greater than the minimum price must have zero mass and Assumption $3$ cannot be satisfied. On the other hand, if the choice probability function is discontinuous at the minimum price at which no one chooses good $1$, then one can always construct an attention distribution $G(\cdot)$ that satisfies Assumption $3$.

For welfare analysis, we consider both the case where Assumption $3$ holds and where it does not. Our finding in this section will be that the distribution of equivalent variation is identified with Assumption $3$ and partially-identified without it.

The following assumption is on the data available to the analyst. We assume rich observed variation in the price of good $1$ holding the price of good $0$ constant.\\

\noindent \textbf{Assumption $4$ (Rich variation in the price of good $1$)}: $q_{1}(p_{1},y)$ is observed for all $p_{1}\in [p_{1}^{\prime},y]$. \\


The interesting setting for welfare analysis of a price increase is when the choice probability of good $1$ is positive at the initial price, $q_{1}(p_{1}^{\prime},y)>0$, because if no one chooses the good at the initial price, then a price increase has no effect on welfare. In this setting, the following empirically defined variable will be relevant for welfare analysis. \\

\noindent \textbf{Definition}: For a given income $y$, let $\overline{p_{1}}$ be the minimum price such that no one chooses good $1$. \\

Given our earlier discussion, we know that $\overline{p_{1}}$ always exists. Finally, we can identify the distribution of equivalent variation for a population with common income $y$. 





\begin{thm} \label{Theorem-1} \textbf{(Welfare Identification}\textbf{)} Suppose Assumptions $1-4$ hold. Consider a price increase from $p_{1}^{\prime}$ to $p_{1}^{\prime\prime}$. Then, the equivalent variation evaluated at income $y$ has a distribution given by 
\[
Pr\{S^{EV}=z\}\\
=\left\{
\begin{array}
[c]{ccc}%
1-q_{1}(p_{1}^{\prime},y) &  & \text{if} \ z=0\\
q_{1}(p_{1}^{\prime},y)-q_{1}(p_{1}^{\prime\prime    },y) &  & \text{if} \ z= \overline{p_{1}}-p_{1}^{\prime}\\
q_{1}(p_{1}^{\prime\prime},y) &  & \text{if} \ z=p_{1}^{\prime\prime}-p_{1}^{\prime}
\end{array}
\right.
\]
Suppose Assumption 3 does not hold. Then, the proportion of consumers who switch from good $1$ to good $0$ in response to the price increase, $q_{1}(p_{1}^{\prime},y)-q_{1}(p_{1}^{\prime\prime    },y)$, have equivalent variation given by
\[
y-p_{1}^{\prime}\geq S^{EV}\geq\overline{p_{1}}-p_{1}^{\prime}.
\]
\end{thm}
Theorem \ref{Theorem-1} point-identifies the distribution of equivalent variation under Assumption 3. Anyone with an equivalent variation of zero must not be choosing good $1$ to begin with. By monotonicity of preference, they will not switch to good $1$ when the price increases. Anyone who has an equivalent variation equal to $\overline{p_{1}}-p_{1}^{\prime}$ must have switched from consuming good $1$ at the original price to consuming  good $0$ at the new price. Thus, the proportion of switchers identifies the proportion with $S^{EV}=\overline{p_{1}}-p_{1}^{\prime}$. Finally, anyone who buys good $1$ at the new price must have also bought the good at the old price (by monotonicity of utility), so their equivalent variation is simply the difference between the two prices. 

Assumption $3$ is necessary for point-identification of EV because it ensures that when the last mass of consumers switches away from good $1$, they are doing so rationally. The price at which these consumers change their choice is $\overline{p_{1}}$. Since they pay full attention at any price, when these consumers change their choice they must be doing so because at that price, they no longer prefer good $1$ to good $0$. In terms of the Bernheim and Rangel (2009) framework, with Assumption $3$, we can observe optimal choices within the "noisy" choice probability function, so welfare analysis is possible. 

Without Assumption $3$, we can only obtain bounds. The upper bound is trivial and comes immediately from Assumption $1(ii)$. The lower bound comes from observing the choices of the consumer with the highest attention threshold. Attention-price thresholds generate ordered consideration sets, where consumers with higher thresholds consider good $1$ at all the prices a consumer with a lower threshold would. Given the common preference, the choice probability only hits zero when either no one prefers good $1$ or the consumer with the highest attention threshold stops paying attention. If this occurs due to attention, then the price at which the consumers would optimally switch is higher, and we have a lower bound for EV. 

An important downside of our welfare identification result is that the analyst draws conclusions about a consumer's welfare that are not based solely on the consumer's own choices. Thus, our welfare identification result is philosophically unappealing compared to the experimental approach, which relies on the consumer's own choices in different experimentally manipulated environments to determine her EV. When possible, one can use the experimental method to corroborate our welfare results.


An interesting observation here is that an EV of $\overline{p_{1}}-p_{1}^{\prime}$ may be greater than or less than the price change. Regardless of the exact value of $\overline{p_{1}}-p_{1}^{\prime}$, the proportion of consumers with equivalent variation equal to it remains the same and can be identified by looking at the proportion of consumers who switch. If we rationalize the choice data using a random utility model (RUM) (Bhattacharya 2015), then the maximum value of the equivalent variation is the price change itself, which is lower than the maximum EV possible with attention-price thresholds. Despite this difference in the possible values of equivalent variation, the type of data required for identification remains the same for binary choice APT models. The analyst only needs to observe choices with variation in the price of good 1 to get the full distribution of EV. With attention-price thresholds, the analyst may require choice data where prices exceed $p_{1}^{\prime\prime}$ to obtain point-identification whereas with RUM, data from choices up to price $p_{1}^{\prime\prime}$ is enough.

\section{\label{Sec-RUMcomparison}Comparison with the Random Utility Model}
In this section, we formally compare the attention-price threshold model and the random utility model. Our main finding on the empirical content of the two models is that any choice data that can be rationalized by a quasi-linear RUM can be rationalized by an APT model. Then, we compare the welfare measures identified under both models and show that they are meaningfully different. 

The binary choice RUM with general unobserved preference heterogeneity is characterized by three axioms (Bhattacharya 2021). We present two of the axioms below for comparison with the APT model: \\

\noindent\textbf{Axiom A-RUM}: $(i) \ q_{1}(p_{1}+\varepsilon,y+\varepsilon)$ is non-increasing in $\varepsilon$ for all $\varepsilon>0$. \newline \indent \indent \indent \indent \indent \vspace{1 mm} $(ii)\ q_{1}(\cdot,y)$ is non-increasing. 

\noindent\textbf{Axiom B-RUM}: $q_{1}(p_{1}+\varepsilon,y+\varepsilon)$ is continuous in $\varepsilon$.\\

First, observe that Axiom A of the APT model and Axiom A-RUM are identical. Second, Axiom B-RUM imposes a continuity requirement on the choice probability function that has no analog in the APT axiomatization. Thus, the APT model is particularly useful in explaining choice data with discontinuities. A final important distinction between the two models is that Axiom B of the APT model places strong restrictions on income effects (see earlier remark on the sign of the income effect) while the RUM places no restrictions on income effects. 
The final RUM axiom (Axiom C in Bhattacharya 2021) characterizes extreme behavior, when the choice probability converges to zero or one. This axiom can be dropped if one imposes a stronger continuity axiom (see online appendix of Bhattacharya 2021).

Given Axiom B, we consider a RUM that also restricts income effects: quasi-linear RUM. Quasi-linear utility models are useful to study binary choice problems when prices are small relative to income (see our discussion in the introduction for references). 

We modify the model from Bhattacharya (2021) to restrict to quasi-linear utility with unobserved preference heterogeneity. Consider our earlier binary choice problem. Assume that the domain of prices and incomes is such that $p_{1}<<y$. Recall that good $0$ has a fixed price of zero and income is common across the population. \\ 

\noindent\textbf{Definition (Quasi-linear RUM)}: A choice probability function $q_{1}(p_{1},y)$ is \textit{Quasi-linear RUM (QRUM) rationalizable} if there exist utility functions 
\[
U_{0}(y,\eta)=V_{0}(\eta)+\beta(\eta)y \ \text{ and } \ U_{1}(y-p_{1},\eta)=V_{1}(\eta)+\beta(\eta)(y-p_{1}) \text{,}
\]
where $\beta(\cdot)>0$, $\eta$ denotes unobserved preference heterogeneity with a distribution $H(\cdot)$, and $V_{0}(\cdot)$ and $V_{1}(\cdot)$ are good-specific components of each consumer's utility, such that 
\begin{align}
    q_{1}(p_{1},y)=\int\mathbb{1}\{V_{0}(\eta)+\beta(\eta)y\leq V_{1}(\eta)+\beta(\eta)(y-p_{1})\}dH(\eta) \text{,} \nonumber \\
\text{equivalently } q_{1}(p_{1})=\int\mathbb{1}\{V_{0}(\eta)\leq V_{1}(\eta)-\beta(\eta)p_{1}\}dH(\eta) \text{,} \label{eqQLRUM}
\end{align}
where $(i)$ for any $p_{1}$, 
\[
\int\mathbb{1}\{V_{0}(\eta)=V_{1}(\eta)-\beta(\eta)p_{1}\}dH(\eta)=0 \text{,}
\]
and $(ii)$ there exists a low price $p_{L}$ such that 
\[
\lim_{p_{1}\searrow p_{L}} Pr[V_{0}(\eta)\leq V_{1}(\eta)-\beta(\eta)p_{1}]=1
\]
\indent \indent \vspace{1 mm} and a high price $p_{H}$ such that \[
\lim_{p_{1}\nearrow p_{H}} Pr[V_{0}(\eta)\leq V_{1}(\eta)-\beta(\eta)p_{1}]=0 \text{.}
\]
The utility from consuming each good depends on a good-specific component and a linear component increasing in the numeraire. We allow unobserved preference heterogeneity in both the good-specific component of utility and the marginal utility of the numeraire. Property $(i)$ states that there are no utility ties, and property $(ii)$ states that if the price is low enough, then everyone prefers good $1$ and if the price is high enough, then no one prefers good $1$. 

We state the characterization of QRUM as a corollary of Theorem 1 in Bhattacharya (2021) and provide a proof in the appendix. The characterization itself is not particularly interesting, but the axioms help us compare the QRUM to the APT model.

The first two axioms of QRUM state that the choice probability is non-increasing and continuous in price. The third and final axiom states that there exists a price low enough such that everyone chooses good $1$, and a price high enough such that no one chooses good $1$. Importantly, the choice probability depends only on price and there are no income effects.

We now present the main result of this section.
\begin{thm}
    \label{QRUM-APT} If a choice probability function $q_{1}(\cdot)$ is QRUM-rationalizable, then it is APT-rationalizable. 
\end{thm}
The QRUM is more restrictive than the APT model in two significant ways: continuity and no income effect. QRUM requires that the choice probability is continuous in price. On the other hand, the APT model does not impose any continuity restrictions on the choice data. The APT model is also more permissive in income effects. The quasi-linear RUM rules out income effects entirely while the APT model allows a restricted form of income effects, as explained earlier.\\

\noindent\textit{Example:} Consider the choice probability function
\begin{align}
    q_{1}(p_{1},y)=\left\{
\begin{array}
[c]{ccc}%
1-\frac{p_{1}}{k} &  & \text{if} \ p_{1}\leq k\\
0 &  & \text{otherwise} \\
\end{array}
\right. \label{example1}
\end{align}
where $k\in(0,y)$. We will show that the choice probability function can be rationalized by both models. Let
\[
U_{0}(y,\eta)=y \text{ and } U_{1}(y-p_{1},\eta)=y-p_{1}+\eta \text{,} 
\]
where $\eta\sim Uniform[0,k]$. Then, the above quasi-linear RUM representation rationalizes the choice probability function. The choice probability satisfies
\begin{align*}
    q_{1}(p_{1},y)=Pr[U_{0}(y,\eta)\leq U_{1}(y-p_{1},\eta)] \\
    =Pr[y\leq y-p_{1}+\eta]=Pr[p_{1}\leq \eta]=1-\frac{p_{1}}{k},
\end{align*}
when $p_{1}\leq k$ and $q_{1}(p_{1},y)=0$, otherwise. For an APT representation, let 
\[
U_{0}(y)=y \text{, } \ U_{1}(y-p_{1})=y-p_{1}+k \text{, and} \ t\sim Uniform[0,k] \text{.}
\]
The choice probability satisfies
\begin{align*}
    q_{1}(p_{1},y)=\left\{
\begin{array}
[c]{ccc}%
1-G(p_{1})=1-\frac{p_{1}}{k} &  & \text{if} \ p_{1}< k\\
0 &  & \text{otherwise.} \\
\end{array}
\right. 
\end{align*}

The choice probability function given by equation (\ref{example1}) admits both a heterogeneous preference explanation, modeled via a quasi-linear RUM, and a heterogeneous attention explanation, modeled via an APT model. In real-world applications, one expects consumers to vary in both their preferences and attention, and researchers would like to identify when one is driving choices versus the other. Theorem \ref{QRUM-APT} can be interpreted as a negative result suggesting that differentiating between the two sources of heterogeneity will be difficult with just choice data. 

Although the QRUM and the APT model can represent the same choice probability function, we show next that the implied welfare conclusions differ starkly. Welfare analysis for the random utility model has been provided by Bhattacharya (2015). Theorem 1 of their paper identifies the distribution of equivalent variation for a binary choice problem. 
In this section, we compare the distributions of equivalent variation identified when the choice data can be rationalized by both models.

We restate here Theorem 1 of Bhattacharya (2015) with minor modifications to match our notation. 
\begin{thm}
\label{thmB2015} \textbf{(Bhattacharya 2015 Theorem 1}\textbf{)} Suppose Assumption 1 holds for each consumer, each consumer can have a different utility function, and all consumers pay attention to both goods at all prices. Data is available according to Assumption 4. Consider a price increase from $p_{1}^{\prime}$ to $p_{1}^{\prime\prime}$. Then, the equivalent variation evaluated at income $y$ has a  distribution given by     
    \[
Pr\{S^{EV}\leq z\}\\
=\left\{
\begin{array}
[c]{ccc}%
0 &  & \text{if} \ z<0\\
1-q_{1}(p_{1}^{\prime}+z,y) &  & \text{if} \ 0\leq z< p_{1}^{\prime\prime}-p_{1}^{\prime}\\
1 &  & \text{if} \ z\geq p_{1}^{\prime\prime}-p_{1}^{\prime}
\end{array}
\right.
\]
\end{thm}
We can compare the distribution of EV identified in Theorem \ref{thmB2015} directly to that identified from Theorem \ref{Theorem-1}. Let $F^{RUM}$ be the distribution of EV identified under RUM, and let $F^{APT}$ be the distribution of EV identified using the APT model.

\begin{thm}
    \label{FOSD} $F^{APT}$ First-order Stochastically Dominates $F^{RUM}$ 
\end{thm}
Both models give the same equivalent variation for consumers who never choose good $1$ and consumers who always choose good $1$. Those who never choose it face no welfare loss from the price increase and those who always choose it face a welfare loss equal to the rise in prices. The two models differ in how they evaluate the welfare change of those who do change their behavior; that is, the equivalent variation of those who chose good $1$ at the original price but choose good $0$ at the new higher price.

RUM assumes that consumers are rational with heterogeneous preferences, so the price of good $1$ at which they substitute away is the price at which they are indifferent between the two goods. Thus, the distribution of EV is a continuum between zero and the price change reflecting the different prices at which each consumer updated their choices. APT takes a different approach: it assumes everyone has the same preference, so everyone is indifferent between the two goods at the same price. Anyone that changes their behavior at any other price is making an error. Under the APT model, this unique indifference price is identified by looking at the behavior of consumers who pay full attention. Given the nature of the consumers' inattention, errors are asymmetric. Limited attention consumers stop paying attention at attention thresholds lower than the indifference price and switch earlier than optimal. Thus, the APT model attributes to all of these consumers the EV of the full attention consumers who switch at the optimal, higher price.\\

\noindent\textit{Example:} Continuing our previous example, consider the choice probability function given by equation (\ref{example1}). As we showed earlier, it can be rationalized by both the QRUM and APT models. We now show that the distributions of EV identified under both are different. 

For a numerical example, let $k=3$. Suppose price increases from $p_{1}^{\prime}=\$1$ to $p_{1}^{\prime\prime}=\$2$. Then, the distribution of EV for the QRUM is given by 
 \[
Pr\{S^{EV}\leq z\}\\
=\left\{
\begin{array}
[c]{ccc}%
0 &  & \text{if} \ z<0\\
\frac{1+z}{3} &  & \text{if} \ 0\leq z< 1\\
1 &  & \text{if} \ z\geq 1 \text{.}
\end{array}
\right.
\]
All consumers who switch from good $1$ to good $0$ when the price increases have an EV strictly less than the price increase, which is of one dollar. Each of those consumers has a different EV corresponding to the price of good $1$ at which they substituted to good $0$. 

Next, we apply Theorem \ref{Theorem-1} to get the distribution of EV under the APT model. Recall that the important price to observe is the minimum price at which no one chooses good $1$, $\overline{p_{1}}$. We can see that for the choice probability function given by equation (\ref{example1}), this minimum price equals $k$. Notice that the choice probability function is continuous at price $k$, so Assumption $3$ cannot be satisfied. Therefore, we can only partially identify the value of EV for those consumers who switch away from good $1$ when the price increases. Applying Theorem \ref{Theorem-1}, we have that those consumers have an EV of at least $2$, which is greater than the price change. Since one-third of the consumers make this switch, one-third of the consumers have an EV of at least $2$. This is strictly greater than the EV assigned to those consumers under QRUM, which was less than the price change. For the remaining consumers we have that
 \[
Pr\{S^{EV}= z\}\\
=\left\{
\begin{array}
[c]{ccc}%
\frac{1}{3} &  & \text{if} \ z=0\\
\frac{1}{3} &  & \text{if} \ z= 1 \text{.}
\end{array}
\right.
\]
Comparing the welfare identification results for the QRUM and APT models, we can see that QRUM underestimates the welfare loss of a price increase compared to the APT model. 

\newpage
\section{Appendix}
\subsection{Equivalent Variation}

\noindent
If the decision maker pays attention to alternative $1$ at both prices (\textit{full attention}), then let $S^{EV}$ be the solution $S$ to the equation
\begin{align}
\max\{U_{0}(y-S),U_{1}(y-S-p_{1}^{\prime})\}  & = 
\max\{U_{0}(y),U_{1}(y-p_{1}^{\prime\prime})\} \label{eq2} 
\end{align}
If the decision maker pays attention to alternative $1$ at only the initial price (\textit{partial attention}), then let $S^{EV}$ be the solution $S$ to the equation
\begin{align}
\max\{U_{0}(y-S),U_{1}(y-S-p_{1}^{\prime})\}  &=U_{0}(y) \label{eq3}
\end{align}
If the decision maker does not pay attention to alternative $1$ at either prices (\textit{no attention}), then let $S^{EV}$ be the solution $S$ to the equation
\begin{align}
U_{0}(y-S)=U_{0}(y) \label{eq4}
\end{align}

\subsection{Proofs for Section \ref{Sec-Characterization}}
\noindent \textbf{Proof of Theorem \ref{Axiomatization}}\\

\noindent \textit{Necessity of Axiom A}\\
First, we show that $q_{1}(p_{1}+\varepsilon,y+\varepsilon)$ is non-increasing in $\varepsilon$ for all $\varepsilon>0$. Let $\varepsilon>0$. By equation (\ref{eq31}), 
\begin{align*}
    q_{1}(p_{1},y)= \mathbb{1}\{U_{0}(y)<U_{1}(y-p_{1})\} \,(1-G(p_{1})) \\
    q_{1}(p_{1}+\varepsilon,y+\varepsilon)= \mathbb{1}\{U_{0}(y+\varepsilon)<U_{1}(y-p_{1})\} \,(1-G(p_{1}+\varepsilon)) 
\end{align*}
Since $U_{0}(\cdot)$ is non-decreasing, $\mathbb{1}\{U_{0}(y+\varepsilon)<U_{1}(y-p_{1})\} \leq \mathbb{1}\{U_{0}(y)<U_{1}(y-p_{1})\}$, and since $G(\cdot)$ is a CDF ($G(\cdot)$ is non-decreasing), $1-G(p_{1}+\varepsilon) \leq 1-G(p_{1})$. Therefore, $q_{1}(p_{1}+\varepsilon,y+\varepsilon) \leq  q_{1}(p_{1},y)$. 

Next, we show that $q_{1}(\cdot,y)$ is non-increasing. Let $p_{1}^{\prime}>p_{1}$. By equation (\ref{eq31}), 
\begin{align*}
    q_{1}(p_{1},y)= \mathbb{1}\{U_{0}(y)<U_{1}(y-p_{1})\} \,(1-G(p_{1})) \\
    q_{1}(p_{1}^{\prime},y)= \mathbb{1}\{U_{0}(y)<U_{1}(y-p_{1}^{\prime})\} \,(1-G(p_{1}^{\prime}))
\end{align*}
Since $y-p_{1}$ is decreasing in $p_{1}$ and $U_{1}(\cdot)$ is increasing, $\mathbb{1}\{U_{0}(y)<U_{1}(y-p_{1}^{\prime})\} \leq \mathbb{1}\{U_{0}(y)<U_{1}(y-p_{1})\}$. $G(\cdot)$ is a CDF, so $1-G(p_{1}^{\prime}) \leq 1-G(p_{1})$. Therefore, $q_{1}(p_{1}^{\prime},y) \leq q_{1}(p_{1},y)$. \\

\noindent \textit{Necessity of Axiom B}\\
Suppose $q_{1}(p_{1},y)>0$ and $q_{1}(p_{1},y^{\prime})>0$ for some $y$ and $y^{\prime}$. By equation (\ref{eq31}), since $q_{1}(p_{1},y)>0$, $\mathbb{1}\{U_{0}(y)<U_{1}(y-p_{1})\}=1$. Similarly, since $q_{1}(p_{1},y^{\prime})>0$, $\mathbb{1}\{U_{0}(y^{\prime})<U_{1}(y^{\prime}-p_{1})\}=1$. Then, $q_{1}(p_{1},y)=1-G(p_{1})=q_{1}(p_{1},y^{\prime})$. \\     

\noindent \textit{Necessity of Axiom C}\\
By Assumption $1(ii)$, for any $y>0$, there exists a price $\bar{p_{1}}\in [0,y]$ such that $U_{0}(y)\geq U_{1}(y-\bar{p_{1}})$. Then, regardless of attention, $q_{1}(\bar{p_{1}},y)=0$. \\

\noindent \textit{Necessity of Axiom D}\\
For a fixed $y>0$, suppose $q_{1}(p_{1}^{\prime},y)=0$ for all $p_{1}^{\prime}>p_{1}$. Assume by contradiction that $q_{1}(p_{1},y)>0$. Since $q_{1}(p_{1},y)>0$, $\mathbb{1}\{U_{0}(y)<U_{1}(y-p_{1})\}=1$ and $1-G(p_{1})>0$. That is, good $1$ is preferred to good $0$, and good $1$ is paid attention to with positive probability. Since $q_{1}(p_{1}^{\prime},y)=0$, it must be that $\mathbb{1}\{U_{0}(y)<U_{1}(y-p_{1}^{\prime})\}=0$ or $1-G(p_{1}^{\prime})=0$ for all $p_{1}^{\prime}>p_{1}$. 

Suppose $1-G(p_{1}^{\prime})=0$ for all $p_{1}^{\prime}>p_{1}$. Since $1-G(p_{1})>0$, there exists a consumer (or continuum of consumers) such that $t_{i}>p_{1}$. Then, for some $p_{1}^{\prime}$ such that $p_{1}<p_{1}^{\prime}<t_{i}$, it must be that $1-G(p_{1}^{\prime})>0$, which is a contradiction. If someone buys good $1$ at price $p_{1}$, then there exists a price $p_{1}^{\prime}$ strictly higher such that this person considers the good at that price as well. In fact, for all $p_{1}^{\prime\prime}$ such that $p_{1}^{\prime\prime} \leq p_{1}^{\prime}$, $1-G(p_{1}^{\prime\prime})>0$. 

For all $p_{1}^{\prime\prime}$ such that $p_{1}<p_{1}^{\prime\prime} \leq p_{1}^{\prime}$, it must be that $\mathbb{1}\{U_{0}(y)<U_{1}(y-p_{1}^{\prime\prime})\}=0$ because $q_{1}(p_{1}^{\prime\prime},y)=0$ for all $p_{1}^{\prime\prime}>p_{1}$. Recall that at price $p_{1}$, $U_{0}(y)<U_{1}(y-p_{1})$ and at price $p_{1}^{\prime}$, $U_{0}(y)\geq U_{1}(y-p_{1}^{\prime})$. By continuity and strict monotonicity of $U_{1}(\cdot)$, there exists a $p_{1}^{\prime\prime}$ such that $p_{1}<p_{1}^{\prime\prime} < p_{1}^{\prime}$ and 
\[
U_{1}(y-p_{1}) > U_{1}(y-p_{1}^{\prime\prime}) > U_{0}(y)\geq U_{1}(y-p_{1}^{\prime}) \text{.}
\]
Contradiction with $\mathbb{1}\{U_{0}(y)<U_{1}(y-p_{1}^{\prime\prime})\}=0$ for all $p_{1}^{\prime\prime}$ such that $p_{1}<p_{1}^{\prime\prime} \leq p_{1}^{\prime}$. 

But if $U_{0}(y)<U_{1}(y-p_{1}^{\prime\prime})$ and $1-G(p_{1}^{\prime\prime})>0$, then $q_{1}(p_{1}^{\prime\prime},y)>0$ for some $p_{1}^{\prime\prime}>p_{1}$. Contradiction with our premise that $q_{1}(p_{1}^{\prime},y)=0$ for all $p_{1}^{\prime}>p_{1}$. \\

\noindent \textit{Necessity of Axiom E}\\
Fix a $y$. Let $p_{1}=0$. By Assumption $2$, everyone pays attention to both goods. Since everyone has the same preference, either everyone buys good $1$ or no one buys good $1$. \\

\noindent \textit{Proof of Sufficiency} \\
First, we modify the notation to aid the proof. Denote by $p>0$ the price of good $1$ and as assumed above, let $p_{0}=0$. Rewrite the choice probability $q(y,y-p) \equiv  q_{1}(p,0,y)$. These two formulations are equivalent since the left-hand side simply represents the choice probability as a function of the residual incomes upon choosing good $0$ or good $1$. Let income $z_{0}\equiv y>0$ and residual income $z_{1}\equiv y-p\in [0,y]$. \\



We restate our model using new notation. Assumption $2$ is as stated above. \\

\noindent\textbf{Assumption $1$}: $(i)$ $U_{0}(z_{0})$ is non-decreasing in $z_{0}$. $U_{1}(z_{1})$ is continuous and strictly increasing in $z_{1}$. \\
\indent $(ii)$ For each $y>0$, there exists a $\underline{z_{1}}(y)\in [0,y]$ such that $U_{0}(y)\geq U_{1}(\underline{z_{1}})$.   \\ 

We restate the axioms on choice behavior in our new notation: \\

\noindent\textbf{Axiom A}: $(i) \ q(\cdot,y-p)$ is non-increasing. $(ii) \ q(y,\cdot)$ is non-decreasing. 

\noindent\textbf{Axiom B}: If $q(z_{0},z_{1})>0$ and $q(z_{0}^{\prime},z_{1}^{\prime})>0$ for $z_{0},z_{1},z_{0}^{\prime},z_{1}^{\prime}$ such that $z_{0}-z_{1}=z_{0}^{\prime}-z_{1}^{\prime}=p$, then
\[
q(z_{0},z_{1})=q(z_{0}^{\prime},z_{1}^{\prime}) \text{.} \
\]
\noindent\textbf{Axiom C}: For any $y>0$, there exists a $\underline{z_{1}}(y)\in [0,y]$ such that $q(y,\underline{z_{1}})=0$. 

\noindent\textbf{Axiom D}: For any $y>0$, if $q(y,z_{1}^{\prime})=0$ for all $z_{1}^{\prime}<z_{1}\in [0,y]$, then $q(y,z_{1})=0$. 

\noindent\textbf{Axiom E}: For any $y>0$, $q(y,y)=0$ or $1$.\\

Axiom A$(i)$ states that holding the residual income fixed, the choice probability of good $1$ is non-increasing in income. Axiom A$(ii)$ states that holding the income fixed, the choice probability of good $1$ is non-decreasing in residual income. Axiom B states that all combinations of income and residual income that have positive choice probability and correspond to the same price have the same choice probability. Axiom C states that for any fixed amount of income, there exists a residual income low enough such that good $1$ is never chosen. For a fixed income, this is equivalent to there being a price high enough such that good $1$ is never chosen. Axiom D states that given a fixed income, if no one chooses good $1$ at any residual income less than $z_{1}$, then no one chooses good $1$ at $z_{1}$ either. Axiom E states that for any income, if residual income equals income, then either everyone chooses good $1$ or no one chooses it.\\ 

We need to construct two utility functions, $U_{0}(\cdot)$ and $U_{1}(\cdot)$, satisfying Assumption $1$, and an attention distribution $G(\cdot)$ satisfying Assumption $2$. \\
    
    By Axiom C, for each $y>0$, there exists a $\underline{z_{1}}(y)\in [0,y]$ such that $q(y,\underline{z_{1}})=0$. Let 
    \[
    U_{0}(y):=\max\{z_{1}:q(y,z_{1})=0\} \ \text{for all} \ y \text{.}
    \]
    That is, $U_{0}(y)$ is the maximum residual income given income $y$ such that no one chooses good $1$. Since the domain for $z_{1}$ is the interval $[0,y]$,  there exists a supremum in $[0,y]$. By Axiom A$(ii)$, $q(y,\cdot)$ is non-decreasing, so the supremum $(z_{1}^{sup})$ is such that for all $z_{1}<z_{1}^{sup}$, $q(y,z_{1})=0$. By Axiom D, 
    $q(y,z_{1}^{sup})=0$, so the 
    maximum is well-defined.\\
    
    First, we show that $U_{0}(y)$ is non-decreasing. Suppose for some $y>y^{\prime}$, $U_{0}(y)<U_{0}(y^{\prime})$. By definition of $U_{0}(y)$,  $q(y,U_{0}(y))=0$. Then, $q(y,U_{0}(y^{\prime}))>0$ because $U_{0}(y)$ is the maximum $z_{1}$ such that $q(y,z_{1})=0$. Since $q(\cdot,U_{0}(y^{\prime}))$ is non-increasing (by Axiom A$(i)$), $q(y^{\prime},U_{0}(y^{\prime}))>0$, contradicting the definition of $U_{0}(y^{\prime})$. Thus,  $U_{0}(y)$ is non-decreasing.\\ 
    
    Next, let 
    \[
    U_{1}(z_{1}):=z_{1}
    \]
    By construction, $U_{1}(z_{1})$ is continuous and increasing, which satisfies Assumption 1$(i)$. \\
    

Finally, we construct the attention distribution $G(t)$. Let 
\[
G(z_{0}-z_{1}):=1-q(z_{0},z_{1}) \text{ for } q(z_{0},z_{1})>0 \text{.}
\]
By Axiom B, $G(\cdot)$ is well-defined in this region. The proportion of consumers with attention thresholds less than or equal to $z_{0}-z_{1}$ is set equal to the proportion of consumers who do not buy good $1$. If $G(\cdot)$ constructed in this way is defined globally, then we are done. By Axiom E, it must be that $G(0)=0$, so Assumption 2 is satisfied.\\

If the $G(\cdot)$ constructed above is not defined for some $t$, then it must be that $q(z_{0},z_{1})=0$ for $t=z_{0}-z_{1}$. Given our construction of utilities, 
\[
q(z_{0},z_{1})=0 \text{ implies } U_{0}(z_{0})\geq U_{1}(z_{1}) \text{.}
\]
Thus, the value of $G(z_{0}-z_{1})$ does not affect choices. If $G(0)$ is not defined, set $G(0)=0$. Then, Assumption $2$ is satisfied. For any remaining $t$ for which $G$ is undefined, extend $G$ so that $G(t)$ is defined. We now have a well-defined CDF $G(\cdot)$. \\

Finally, we show that the utility functions and attention distribution above satisfy the rationalizability condition (\ref{eq31})
\[
q(y,y-p)=\mathbb{1}\{U_{0}(y)<U_{1}(y-p)\} \,(1-G(p)) \text{.}
\]
Notice that 
\[
q(z_{0},z_{1})=0 \ \text{iff} \ U_{0}(z_{0})\geq U_{1}(z_{1}) \text{, and}
\]
\[
q(z_{0},z_{1})=v>0 \ \text{iff} \ U_{0}(z_{0})< U_{1}(z_{1}) \ \text{and} \ 1-G(z_{0}-z_{1})=v
\]
Finally, it is easy to check that for each $y>0$, there exists a $\underline{z_{1}}(y)\in [0,y]$ such that $U_{0}(y)\geq U_{1}(\underline{z_{1}})$, implying Assumption $1(ii)$. 
{\normalsize \hfill
$\blacksquare$}
\subsection{Proofs for Section \ref{Sec-Welfare}}
The proof of Theorem \ref{Theorem-1} (Welfare Identification) requires obtaining all the possible values of equivalent variation (EV). We present these in Lemma \ref{lemma-1} below. To cleanly state some of the values of EV, it is helpful to first define the price of good $1$ such that consumers are indifferent between good $1$ and good $0$.

Formally, let $p^{10}$ be the price of good $1$ such that 
\[
U_{1}(y-p^{10})=U_{0}(y) \text{.}
\]
This price can be understood as the \textit{reservation price} for rational consumers, who do not have limited attention.

\begin{lemma}
\label{lemma-1} Consider a binary choice setting with good $0$ and good $1$. Suppose Assumptions $1$ and $2$ hold. The price of good $1$ increases from $p_{1}^{\prime}$ to $p_{1}^{\prime\prime}$. The equivalent variation for each consumer is as follows: 
\begin{enumerate}
    \item If $p_{1}^{\prime}\geq t_{i}$ (no attention), then regardless of the preference, $S^{EV}=0$. 
    \item If  $U_{1}(y-p_{1}^{\prime})\leq U_{0}(y)$, then regardless of the consumer's attention, $S^{EV}=0$. 
    \item If $U_{1}(y-p_{1}^{\prime\prime})\leq U_{0}(y)<U_{1}(y-p_{1}^{\prime})$, then for both full attention ($p_{1}^{\prime}<p_{1}^{\prime\prime}<t_{i}$) and partial attention ($p_{1}^{\prime}<t_{i}\leq p_{1}^{\prime\prime}$) consumers, \[ S^{EV}=y-p_{1}^{\prime}-U_{1}^{-1}(U_{0}(y))=p^{10}-p_{1}^{\prime}\leq p_{1}^{\prime\prime}-p_{1}^{\prime}.\]
    \item  If $U_{0}(y)<U_{1}(y-p_{1}^{\prime\prime})<U_{1}(y-p_{1}^{\prime})$,
    \begin{enumerate}
        \item and $p_{1}^{\prime}<p_{1}^{\prime\prime}<t_{i}$ (full attention), then $S^{EV}=p_{1}^{\prime\prime}-p_{1}^{\prime}$.
        \item and $p_{1}^{\prime}<t_{i}\leq p_{1}^{\prime\prime}$ (partial attention), then \[S^{EV}=y-p_{1}^{\prime}-U_{1}^{-1}(U_{0}(y))=p^{10}-p_{1}^{\prime}>p_{1}^{\prime\prime}-p_{1}^{\prime} .\] 
    \end{enumerate}
\end{enumerate}
\end{lemma}

\noindent\textbf{Proof of Lemma \ref{lemma-1}} \\
\indent \textbf{Case 1}: Suppose $p_{1}^{\prime}>t_{i}$ (No attention). \\
Then, equation (\ref{eq4}) gives 
\[
U_{0}(y-S)=U_{0}(y) \text{.}
\]
If $S>0$, then by monotonicity, $U_{0}(y-S)<U_{0}(y)$. Thus, $S^{EV}=0$. \\

\indent \textbf{Case 2}: Suppose $U_{1}(y-p_{1}^{\prime})\leq U_{0}(y)$. \\
For a \textit{full attention consumer}, equation (\ref{eq2}) gives us
\[
\max\{U_{0}(y-S),U_{1}(y-S-p_{1}^{\prime})\} = 
\max\{U_{0}(y),U_{1}(y-p_{1}^{\prime\prime})\}.
\]
If $S>0$, then
\begin{align*}
    \max\{U_{0}(y-S),U_{1}(y-S-p_{1}^{\prime})\} < 
\max\{U_{0}(y),U_{1}(y-p_{1}^{\prime})\} &\\
=U_{0}(y) &\\
=\max\{U_{0}(y),U_{1}(y-p_{1}^{\prime\prime})\}. 
\end{align*}
This contradicts equation (\ref{eq2}), so $S^{EV}=0$. \\
For a \textit{partial attention consumer}, equation (\ref{eq3}) gives us
\[
\max\{U_{0}(y-S),U_{1}(y-S-p_{1}^{\prime})\} = 
U_{0}(y).
\]
If $S>0$, then
\begin{align*}
    \max\{U_{0}(y-S),U_{1}(y-S-p_{1}^{\prime})\} < 
\max\{U_{0}(y),U_{1}(y-p_{1}^{\prime})\} &\\
=U_{0}(y).
\end{align*}
This contradicts with equation (\ref{eq3}), so $S^{EV}=0$. \\

\indent \textbf{Case 3}: Suppose $U_{1}(y-p_{1}^{\prime\prime})\leq U_{0}(y)<U_{1}(y-p_{1}^{\prime})$.\\
For a \textit{full attention consumer}, equation (\ref{eq2}) gives us
\[
\max\{U_{0}(y-S),U_{1}(y-S-p_{1}^{\prime})\} = 
\max\{U_{0}(y),U_{1}(y-p_{1}^{\prime\prime})\}.
\]
Given our preference, the RHS equals $U_{0}(y)$. Thus,
\begin{align}
    \max\{U_{0}(y-S),U_{1}(y-S-p_{1}^{\prime})\} =U_{0}(y) \text{.}\label{eq51}
\end{align}
Suppose $U_{1}(y-S-p_{1}^{\prime})\leq U_{0}(y-S)$. Then, equation (\ref{eq51}) becomes 
\[
U_{0}(y-S)=U_{0}(y),
\]
 and $S=0$ by strict monotonicity of $U_{0}$. But then,
\[
 U_{0}(y-S)=U_{0}(y)>U_{1}(y-S-p_{1}^{\prime}) =  U_{1}(y-p_{1}^{\prime}) \text{.}
 \]
This contradicts our initial assumption that $U_{0}(y)<U_{1}(y-p_{1}^{\prime})$. Thus, $U_{1}(y-S-p_{1}^{\prime})>U_{0}(y-S)$. Then, equation (\ref{eq51}) gives
 \[
 U_{1}(y-S-p_{1}^{\prime})=U_{0}(y) \text{.}
 \]
Taking the inverse and rearranging gives 
\[
S^{EV}=y-p_{1}^{\prime}-U_{1}^{-1}(U_{0}(y)) \text{.}
\]
Observe that 
\[
 U_{1}(y-(S^{EV}+p_{1}^{\prime}))=U_{0}(y) \text{,}
\]
where $p^{10}=S^{EV}+p_{1}^{\prime}$.
Finally, notice that since $U_{1}(y-p_{1}^{\prime\prime})\leq U_{0}(y)$, $S^{EV}\leq p_{1}^{\prime\prime}-p_{1}^{\prime}$. \\
For a \textit{partial attention consumer}, equation (\ref{eq3}) gives us
\[
\max\{U_{0}(y-S),U_{1}(y-S-p_{1}^{\prime})\} = 
U_{0}(y).
\]
This is equivalent to equation  (\ref{eq51}), and the rest of the argument follows as for the full attention consumer. \\

\indent \textbf{Case 4}: Suppose $U_{0}(y)<U_{1}(y-p_{1}^{\prime\prime})<U_{1}(y-p_{1}^{\prime})$. \\ 
For a \textit{full attention consumer}, the proof is exactly as follows from Bhattacharya (2015). 
For a \textit{partial attention consumer}, the proof is identical to the proof for the partial attention consumer in Case 3, except for the last line. Notice now that since $U_{1}(y-p_{1}^{\prime\prime})>U_{0}(y)$, $S^{EV}>p_{1}^{\prime\prime}-p_{1}^{\prime}$. {\normalsize \hfill
$\blacksquare$} \\

Lemma \ref{lemma-1} states that there are three possible values for EV: $0$, $p_{1}^{\prime\prime}-p_{1}^{\prime}$, and $p^{10}-p_{1}^{\prime}$. To identify the last value of EV, it is necessary to identify $p^{10}$, the price of good 1 at which the consumer is indifferent between good $1$ and good 0 (recall we break utility ties in favor of good $0$). However, this indifference price is only of interest if consumers switch from good $1$ to good $0$ when the price of good $1$ changes. A prerequisite for that is some consumers must choose good $1$ at the initial price; that is, $q_{1}(p_{1}^{\prime},y)>0$. If good $1$ is chosen at the initial price, Assumption $1$ implies that, for any income, there exists a unique price $p^{10}$.
We state this formally as an observation. \\

\noindent \textbf{Observation}: For each $y>0$, if $q_{1}(p_{1}^{\prime},y)>0$, then there exists a unique $p^{10}$ such that $U_{1}(y-p^{10})=U_{0}(y)$. \\

Next, we show that $p^{10}$ can be identified from choice data. The following lemma shows that when good $1$ is chosen at the initial price, the empirically observed price $\overline{p_{1}}$ is a lower bound of $p^{10}$. If we also impose Assumption $3$, then $\overline{p_{1}}$ is in fact $p^{10}$. That is, given our data, $p^{10}$ is identified.

\begin{lemma}
\label{lemma-2} If good $1$ is chosen at the initial price, then $\overline{p_{1}}\leq p^{10}$. In addition, if Assumption $3$ holds, then $\overline{p_{1}}= p^{10}$
\end{lemma}
\noindent \textbf{Proof of Lemma} \textbf{\ref{lemma-2} } \\
Since good $1$ is chosen at the initial price, we know that a unique $p^{10}$ exists. First, we show that $\overline{p_{1}} \leq p^{10}$. For $p_{1}\geq p^{10}$,  $U_{1}(y-p_{1})\leq U_{0}(y)$ by monotonicity. Good $0$ is always considered, so no one chooses good $1$. Thus, $q_{1}(p_{1},y)=0$ for all $p_{1}\geq p^{10}$. Therefore, $\overline{p_{1}} \leq p^{10}$. \\

Now, we show that $\overline{p_{1}} \geq p^{10}$. For $p_{1} < p^{10}$, $U_{1}(y-p_{1}) > U_{0}(y)$ by monotonicity. Given Assumption $3$, there exists someone who pays attention to good $1$ for all $p_{1} < p^{10}$, thus $q_{1}(p_{1},y)>0$ for all $p_{1} < p^{10}$. Therefore, $\overline{p_{1}} \geq p^{10}$. {\normalsize \hfill
$\blacksquare$}\\

When consumers are indifferent between the two goods, no one chooses good $1$, so the minimum price such that no one chooses good $1$ must be less than or equal to the price at which consumers are indifferent, which gives us our lower bound. Assumption $3$ states that for any given price of good $1$ there exist consumers who pay attention. These consumers would choose good $1$ as long as good $1$ at its price was preferable to good $0$. Thus, the minimum price such that no one chooses good $1$ must be greater than or equal to the price at which consumers are indifferent between good $1$ and good $0$. \\

\noindent \textbf{Proof of Theorem \ref{Theorem-1}}\\
First, we determine the probability that $S^{EV}=0$. Lemma \ref{lemma-1} states that this occurs when individuals do not pay attention to good $1$ at the original price or they prefer good $0$ to good $1$. 
\[
Pr\{S^{EV}=0\}\\
=Pr\{p_{1}^{\prime}\geq t_{i} \ \text{or} \ U_{1}(y-p_{1}^{\prime})\leq U_{0}(y)\}
\]
Any consumer that does not pay attention to good $1$ at the price $p_{1}^{\prime}$ does not buy good $1$. Any consumer that prefers good $0$ over good $1$ at $p_{1}^{\prime}$ will not buy good $1$ regardless of whether they consider good $1$. Thus, both types of consumers will not buy good $1$ at the original price. Conversely, consider a consumer who does not buy good $1$ at the price $p_{1}^{\prime}$. If she paid attention to good $1$, and preferred it to good $0$, then she would buy good $1$. Thus, if she does not buy good $1$, then she must not pay attention to good $1$ or she prefers good $0$ to good $1$. Therefore,
\begin{align*}
    Pr\{S^{EV}=0\}
 =Pr\{p_{1}^{\prime}\geq t_{i} \ \text{or} \ U_{1}(y-p_{1}^{\prime})\leq U_{0}(y)\} = 1-q_{1}(p_{1}^{\prime},y) \text{.}
\end{align*}
Next, we determine the probability that $S^{EV}=p_{1}^{\prime\prime}-p_{1}^{\prime}$. Lemma \ref{lemma-1} states that this occurs for individuals who pay attention to good $1$ at both the old and new prices, and prefer good $1$ at $p_{1}^{\prime\prime}$ to good $0$. That is, 
\begin{align*}
    Pr\{S^{EV}=p_{1}^{\prime\prime}-p_{1}^{\prime}\}
=Pr\{U_{0}(y)<U_{1}(y-p_{1}^{\prime\prime})<U_{1}(y-p_{1}^{\prime}) \ \text{and} \ p_{1}^{\prime}<p_{1}^{\prime\prime}<t_{i}\} \text{.}
\end{align*}
Anyone that prefers good $1$ at the price $p_{1}^{\prime\prime}$ to good $0$, and pays attention to it will buy good $1$. Conversely, anyone who buys good $1$ at price $p_{1}^{\prime\prime}$ must prefer good $1$ to good $0$ and must pay attention to good $1$. Therefore,
\begin{align*}
    Pr\{S^{EV}=p_{1}^{\prime\prime}-p_{1}^{\prime}\}
=Pr\{U_{0}(y)<U_{1}(y-p_{1}^{\prime\prime})<U_{1}(y-p_{1}^{\prime}) \ \text{and} \ p_{1}^{\prime}<p_{1}^{\prime\prime}<t_{i}\} \\=q_{1}(p_{1}^{\prime\prime},y) \text{.}
\end{align*}
Finally, we determine the probability that $S^{EV}=\overline{p_{1}}-p_{1}^{\prime}$. By Lemma \ref{lemma-1}, these are all the remaining types of individuals. 
\begin{align*}
    Pr\{S^{EV}=\overline{p_{1}}-p_{1}^{\prime}\}\\
=Pr\{U_{1}(y-p_{1}^{\prime\prime})\leq U_{0}(y)<U_{1}(y-p_{1}^{\prime}) \ \text{and} \ (p_{1}^{\prime}<p_{1}^{\prime\prime}<t_{i} \ \text{or} \ p_{1}^{\prime}<t_{i}\leq p_{1}^{\prime\prime} )\} \\ 
+ Pr\{U_{0}(y)<U_{1}(y-p_{1}^{\prime\prime})<U_{1}(y-p_{1}^{\prime}) \ \text{and} \ p_{1}^{\prime}<t_{i}\leq p_{1}^{\prime\prime}\} \text{.}
\end{align*}
Under both preferences and possible attention thresholds, the consumer pays attention to good $1$ at the original price, and prefers good $1$ to good $0$, so she must buy good $1$. When the price of good $1$ increases to $p_{1}^{\prime\prime}$, the consumer who continues to prefer good $1$ at the new price to good $0$ does not pay attention to good $1$ at $p_{1}^{\prime\prime}$, so she cannot buy good $1$. The consumer with the other preference does not prefer good $1$ at the new price to good $0$, so she also does not buy good 1. Consumers of these types switch from buying good 1 at the old price to not buying good 1 at the new price. Conversely, suppose a consumer does buy good $1$ at $p_{1}^{\prime}$ but does not buy good $1$ at $p_{1}^{\prime\prime}$. Then, the consumer must pay attention to good $1$ at the price $p_{1}^{\prime}$ and prefer good $1$ to good $0$. Furthermore, the consumer must not prefer good $1$ at the price $p_{1}^{\prime\prime}$ or must not pay attention to good $1$ at the price $p_{1}^{\prime\prime}$, otherwise the consumer would buy good $1$ at the new price. These consumers are exactly those with the above utility functions and attention thresholds. Therefore,
\begin{align*}
    Pr\{S^{EV}=\overline{p_{1}}-p_{1}^{\prime}\}\\
=Pr\{U_{1}(y-p_{1}^{\prime\prime})\leq U_{0}(y)<U_{1}(y-p_{1}^{\prime}) \ \text{and} \ (p_{1}^{\prime}<p_{1}^{\prime\prime}<t_{i} \ \text{or} \ p_{1}^{\prime}<t_{i}\leq p_{1}^{\prime\prime} )\} \\ 
+ Pr\{U_{0}(y)<U_{1}(y-p_{1}^{\prime\prime})<U_{1}(y-p_{1}^{\prime}) \ \text{and} \ p_{1}^{\prime}<t_{i}\leq p_{1}^{\prime\prime}\} \\ = q_{1}(p_{1}^{\prime},y)-q_{1}(p_{1}^{\prime\prime},y) \text{.}
\end{align*}
{\normalsize \hfill
$\blacksquare$}
\subsection{Proofs for Section \ref{Sec-RUMcomparison}}
We provide the empirical content of binary choice QRUM below. It is characterized by three axioms.\\

\noindent\textbf{Axiom A-QRUM}: $q_{1}(\cdot)$ is non-increasing. \\
\noindent\textbf{Axiom B-QRUM}: $q_{1}(\cdot)$ is continuous.\\
\noindent\textbf{Axiom C-QRUM}: There exists a low price $p_{L}$ and a high price $p_{H}$ such that 
\[
\lim_{p_{1}\searrow p_{L}}q_{1}(p_{1})=1 \text{ and }\lim_{p_{1}\nearrow p_{H}}q_{1}(p_{1})=0 \text{,}
\]
respectively.

We state the characterization of QRUM as a corollary of Theorem 1 in Bhattacharya (2021).
\begin{corol}
    \label{QRUM-Axiomatization} A choice probability function $q_{1}(\cdot)$ is QRUM-rationalizable if and only if it satisfies Axioms (A-C)-QRUM. 
\end{corol}

\noindent \textbf{Proof of Corollary \ref{QRUM-Axiomatization}}\\
We follow the proof of Bhattacharya 2021 Theorem 1 with appropriate modifications for the quasi-linear utility model.\\ 
\textit{Proof of Necessity}\\
The rationalizability condition for QRUM is given by equation (\ref{eqQLRUM}):
\[
 q_{1}(p_{1})=\int\mathbb{1}\{V_{0}(\eta)\leq V_{1}(\eta)-\beta(\eta)p_{1}\}dH(\eta)\text{.}
\]
Rearranging, we get
\[
 q_{1}(p_{1})=\int\mathbb{1}\{\frac{V_{1}(\eta)-V_{0}(\eta)}{\beta(\eta)}\geq p_{1}\} dH(\eta) \text{.}
\]
Since $\frac{V_{1}(\eta)-V_{0}(\eta)}{\beta(\eta)}$ is a function only of $\eta$, let 
\[
f(\eta):=\frac{V_{1}(\eta)-V_{0}(\eta)}{\beta(\eta)} \text{.}
\]
Then, we can rewrite the rationalizability condition as
\begin{align}
    q_{1}(p_{1})=\int\mathbb{1}\{f(\eta)\geq p_{1}\} dH(\eta) \label{eqQRUMproof} \text{.}
\end{align}
It is simple to check that Axioms (A-C)-QRUM are satisfied. \\

\noindent\textit{Proof of Sufficiency}\\
By Axiom C-QRUM, for any $\nu \in [0,1]$, the set $\{p_{1}:q_{1}(p_{1})\geq\nu\}$ is non-empty. For any $\nu \in [0,1]$, define
\begin{align}
    q_{1}^{-1}(\nu):=sup\{p_{1}:q_{1}(p_{1})\geq\nu\} \label{QRUM-quantile} \text{,}
\end{align}
which takes values in $[0,y)$, given our domain for prices is such that $p_{1}<<y$. By Axiom A-QRUM, $q_{1}^{-1}(\cdot)$ is non-increasing. 

Now consider a random variable $N\sim Uniform(0,1)$. Define
\[
f(N):= q_{1}^{-1}(N) \text{.}
\]
Observe that the function $ 1-q_{1}(\cdot)$ is a continuous CDF (by Axioms (A-C)-QRUM). By definition, $q_{1}^{-1}(\nu)$ is the $(1-\nu)^{th}$ quantile. 

Consider the following properties of quantiles (see Appendix B in Bhattacharya 2021 for more general statements and proofs of the properties):
\begin{enumerate}
    \item For any $\nu \in [0,1]$, it must be that $q_{1}(q_{1}^{-1}(\nu))=\nu$.
    \item For any $\nu \in [0,1]$ and $p_{1}$, we have that $q_{1}(p_{1})\geq \nu$.
    \item The function $q_{1}^{-1}(\cdot)$ is one-to-one on $[0,1]$.
\end{enumerate}
Then, we have that 
\begin{align}
    Pr(q_{1}^{-1}(N)\geq p_{1})=Pr(N\leq q_{1}(p_{1}))=q_{1}(p_{1}) \label{eqQRUM3} \text{,}
\end{align}
where the first equality follows from property $(2)$ and the second equality follows from the uniform distribution of $N$. 

We have constructed $f(N)$ that satisfies the rationalizability condition (\ref{eqQRUMproof}). Construct the following three functions from $f(N)$: $V_{0}(N), V_{1}(N),$ and $\beta(N)$. The only restrictions they must satisfy are that $\beta(N)>0$ and
\[
f(N)=\frac{V_{1}(N)-V_{0}(N)}{\beta(N)} \text{.}
\]
Finally, we can construct our utility functions so that 
\[
U_{0}(y,N):=V_{0}(N)+\beta(N)y \ \text{ and } \ U_{1}(y-p_{1},N):=V_{1}(N)+\beta(N)(y-p_{1}) \text{,}
\]
which satisfy our quasi-linear utility definition.

Next, for any $\nu,\nu^{\prime} \in [0,1]$ such that $\nu \neq \nu^{\prime}$, we cannot have that $q_{1}^{-1}(\nu)=q_{1}^{-1}(\nu^{\prime})$ by property $(3)$. Therefore, for all $p_{1}$
\[
Pr[q_{1}^{-1}(N)= p_{1}]=0 \text{,}
\]
which implies condition $(i)$ of "no utility ties" in our QRUM definition. 

Finally, we have
\[
 \lim_{p_{1}\searrow p_{L}}Pr[q_{1}^{-1}(N)\geq  p_{1}] = \lim_{p_{1}\searrow p_{L}}Pr[N\leq q_{1}(p_{1})] = \lim_{p_{1}\searrow p_{L}}q_{1}(p_{1})=1 \text{,}
\]
where the first two equalities follow from equation (\ref{eqQRUM3}), and the final equality is implied by Axiom C-QRUM. Similarly, we can show that 
\[
\lim_{p_{1}\nearrow p_{H}}Pr[q_{1}^{-1}(N)\geq  p_{1}] =0 \text{,}
\]
which implies condition $(ii)$ of our QRUM definition. {\normalsize \hfill
$\blacksquare$}\\

\noindent \textbf{Proof of Theorem \ref{QRUM-APT}}\\
Suppose $q_{1}(p_{1},y)$ is quasi-linear RUM rationalizable. Then, by Corollary \ref{QRUM-Axiomatization}, it satisfies Axioms (A-C)-QRUM. We now show that it must satisfy Axioms A-E of the APT model. 

Axiom A follows immediately from Axiom A-QRUM since for any quasi-linear RUM rationalizable choice probability function $q_{1}(p_{1},y)=q_{1}(p_{1})$. Since $q_{1}(\cdot)$ is non-increasing, we have that $q_{1}(p_{1}+\varepsilon,y+\varepsilon)$ is also non-increasing in $\varepsilon$ for all $\varepsilon>0$, which satisfies Axiom A.

Axiom B also follows from Axiom A-QRUM because for any $y,y^{\prime}$ and $p_{1}$, if $q_{1}(p_{1},y)>0$ and $q_{1}(p_{1},y^{\prime})>0$, then
\[
q_{1}(p_{1},y) =q_{1}(p_{1})=q_{1}(p_{1},y^{\prime}) \text.
\]
\indent Given the domain $p_{1}<<y$ and Axiom C-QRUM, price $p_{H}$ satisfies $q_{1}(p_{H},y)=0$ for all $y$, which implies Axiom C. 

Suppose for any $y$, $q_{1}(p_{1}^{\prime},y)=0$ for all $p_{1}^{\prime}>p_{1}$. We know that $q_{1}(p_{1}^{\prime},y)=q_{1}(p_{1}^{\prime})$ since the choice probability function is QRUM-rationalizable. By Axiom B-QRUM, $q_{1}(\cdot)$ is continuous, so $q_{1}(p_{1},y)=0$, which implies Axiom D. 

Finally, Axiom C-QRUM states that there exists a low price $p_{L}$ such that $\lim_{p_{1}\searrow p_{L}}q_{1}(p_{1})=1 $. Then, $q_{1}(0,y)=1$, which satisfies Axiom E. {\normalsize \hfill
$\blacksquare$}\\

\noindent \textbf{Proof of Theorem \ref{FOSD}}\\
Observe that $F^{APT}(0)=F^{RUM}( 0)=1-q_{1}(p_{1}^{\prime},y)$. In Theorem \ref{Theorem-1}, only one of the following can be true: either $\overline{p_{1}}-p_{1}^{\prime}\leq p_{1}^{\prime\prime}-p_{1}^{\prime}$ or $\overline{p_{1}}-p_{1}^{\prime}>p_{1}^{\prime\prime}-p_{1}^{\prime}$. \\

\noindent \textit{First case}: suppose $\overline{p_{1}}-p_{1}^{\prime}\leq p_{1}^{\prime\prime}-p_{1}^{\prime}$.
If $z=\overline{p_{1}}-p_{1}^{\prime}$, then $q_{1}(p_{1}^{\prime}+z,y)=0$ and $q_{1}(p_{1}^{\prime}+z+\epsilon,y)=0$ for all $\epsilon>0$ by the definition of $\overline{p_{1}}$. Thus, 
\[
F^{APT}(z)=F^{RUM}( z)=1-q_{1}(p_{1}^{\prime}+z,y)=1 \text{.}
\]
For all $0<z<\overline{p_{1}}-p_{1}^{\prime}$, 
\begin{align*}
    F^{APT}(z)=1-q_{1}(p_{1}^{\prime},y) \text{,}
\end{align*}
since the only possible value of equivalent variation $S^{EV}<\overline{p_{1}}-p_{1}^{\prime}\leq p_{1}^{\prime\prime}-p_{1}^{\prime}$ in the APT model is zero. By Theorem \ref{thmB2015}, we have that 
\begin{align*}
F^{RUM}(z)= 1-q_{1}(p_{1}^{\prime}+z,y) \text{.}
\end{align*}
Since $q_{1}(p_{1},y)$ is non-increasing in $p_{1}$, we have that $F^{RUM}(z)\geq F^{APT}(z)$. Thus, $F^{APT}$ first-order stochastically dominates $F^{RUM}$.\\

\noindent \textit{Second case}: suppose $\overline{p_{1}}-p_{1}^{\prime}>p_{1}^{\prime\prime}-p_{1}^{\prime}$. For $z< p_{1}^{\prime\prime}-p_{1}^{\prime}$, we get that
\begin{align*}
     F^{APT}(z)=1-q_{1}(p_{1}^{\prime},y)\leq 1-q_{1}(p_{1}^{\prime}+z,y)=F^{RUM}(z) \text{.}
\end{align*}
For $z= p_{1}^{\prime\prime}-p_{1}^{\prime}$, we get 
\begin{align*}
    F^{APT}(z)=1-q_{1}(p_{1}^{\prime},y)+q_{1}(p_{1}^{\prime\prime},y)\leq  1=F^{RUM}(z)
\end{align*}
Finally, observe that $z>p_{1}^{\prime\prime}-p_{1}^{\prime}$ is not in the support of $F^{RUM}$ while it is in the support of $F^{APT}$. Therefore, $F^{APT}$ first-order stochastically dominates $F^{RUM}$ {\normalsize \hfill
$\blacksquare$}

\end{document}